\documentclass[preprint,12pt,table]{elsarticle}

\usepackage{amssymb}
\usepackage[dvipsnames]{xcolor}
\usepackage{tabularx}
\usepackage{longtable}
\usepackage{multirow}
\usepackage{amssymb}% http://ctan.org/pkg/amssymb
\usepackage{pifont}% http://ctan.org/pkg/pifont
\newcommand{\cmark}{\ding{51}}%
\newcommand{\xmark}{\ding{55}}%

\usepackage{hyperref}
\usepackage{soul}

\journal{Environmental modelling \& software}

\begin{document}
\begin{frontmatter}



\title{A Collaborative Platform for Soil Organic Carbon Inference Based on Spatiotemporal Remote Sensing Data}


\author[inst1]{Jose Manuel Aroca-Fernandez}

\affiliation[inst1]{organization={Departamento de Ingenieria Informatica, Universidad de Burgos},%Department and Organization
            addressline={Avda. Cantabria s/n},
            city={Burgos},
            postcode={09006},
            country={Spain}}

\affiliation[inst2]{organization={School of Mathematics, University of Edinburgh},%Department and Organization
            addressline={James Clerk Maxwell Building, Peter Guthrie Tait Rd},
            city={Edinburgh},
            postcode={EH9 3FD},
            country={United Kingdom}}

\affiliation[inst3]{organization={Image Processing Laboratory (IPL), E4 building - 4th floor, Parc Cientific Universitat de Valencia},%Department and Organization
            addressline={C/ Cat. Agustin Escardino Benlloch, 9},
            city={Valencia},
            postcode={46980},
            country={Spain}}

\author[inst1]{Jose Francisco Diez-Pastor}
\author[inst1]{Pedro Latorre-Carmona}
\author[inst2]{Victor Elvira}
\author[inst3]{Gustau Camps-Valls}
\author[inst1]{Rodrigo Pascual}
\author[inst1]{Cesar Garcia-Osorio}

\begin{abstract}
Soil organic carbon (SOC) is a key indicator of soil health, fertility, and carbon sequestration, making it essential for sustainable land management and climate change mitigation. However, large-scale SOC monitoring remains challenging due to spatial variability, temporal dynamics, and multiple influencing factors.
We present WALGREEN, a platform that enhances SOC inference by overcoming  limitations of current applications. Leveraging machine learning and diverse soil samples, WALGREEN generates predictive models using historical public and private data.
Built on cloud-based technologies, it offers a user-friendly interface for researchers, policymakers, and land managers to access carbon data, analyze trends, and support evidence-based decision-making. Implemented in Python, Java, and JavaScript, WALGREEN integrates Google Earth Engine and Sentinel Copernicus via scripting, OpenLayers, and Thymeleaf in a Model-View-Controller framework. This paper aims to advance soil science, promote sustainable agriculture, and drive critical ecosystem responses to climate change.

\end{abstract}

%    \centering
%    \includegraphics[width=1\linewidth]{images/Platformmainview1.png}
%    \label{fig:abstract}
    %\caption{Platform main view}

%\item The need to democratize and simplify technical access to satellite imagery is increasing.
%\item Numerous public datasets with onsite measurements of soil organic carbon are available.
%\item Access to and integration of TIFF images, satellite data, and public datasets are essential.
%\item An easy-to-use and free platform is developed
%\item System integration is crucial, but deep programming knowledge is not required.
%\item Applicability and functionality are core aims of the platform.



\end{frontmatter}



%%%%%%%%%%%%%%%%%%%%%%%%%%%%%%%%%%%%%%%%%%%%%%%%%%%%%%%%%%%%%%%%%%%%%%%%%%%%%%%%%%%
\section{Introduction}
%%%%%%%%%%%%%%%%%%%%%%%%%%%%%%%%%%%%%%%%%%%%%%%%%%%%%%%%%%%%%%%%%%%%%%%%%%%%%%%%%%%
\label{sec:intro}
Soil organic carbon (SOC) is a key factor in soil health, contributing to processes governing carbon cycling, agricultural productivity, climate change mitigation, and food security~\cite{Scharlemann201481} on a global scale. SOC is vital for sustainable land management and environmental conservation. Therefore, its estimation and monitoring are crucial.

The ``4 per mile Soils for Food Security and Climate" initiative was launched at the Conference of the Parties (COP21)~\cite{robbins2016understand} with the goal of increasing global soil organic matter stocks by 4 per 1000 (or $0.4\%$) per year to compensate for the global emissions of greenhouse gases by anthropogenic sources. SOC storage offers the possibility to reduce the concentration of carbon dioxide in the atmosphere~\cite{toth2013lucas}.

Conventional SOC quantification approaches (e.g., soil sampling and laboratory analysis) are labor-intensive, time-consuming, and expensive. However, in recent years, SOC inference has significantly mitigated these challenges  owing to several improvements, including remote sensing techniques, machine learning (ML) algorithms, and collaborative data sharing.

In this paper, we propose a new application called WALGREEN\footnote{https://walgreen.ngrok.io/} (Web-based Analytical Learning for Green Environments). This collaborative platform aims to infer and map SOC from remote sensing and in situ data. ML algorithms combine these information sources (remote sensing images, soil surveys, geospatial data) to infer SOC. WALGREEN helps bridge the gap between scientific research and real-world application by providing a comprehensive and user-friendly interface for researchers, policymakers, and agricultural stakeholders. The collaborative nature of WALGREEN allows users to contribute and access high-quality SOC data.

Remote sensing offers solutions for estimating SOC in all types of soils and ecosystems that will assist policymakers, land managers, and field researchers in monitoring and managing soil health and carbon sequestration.

Remote sensing and onsite data integration refers to combining remote sensing data obtained from satellite, aerial, or other sensor-based platforms with data collected directly from the field or onsite. This integrated approach allows for a more complete and accurate understanding of environmental and agricultural systems by combining the strengths of both remote (spatial and broad) and onsite (local and specific) data. WALLGREEN is a fully functional remote sensing platform that has been developed on four pillars:
\begin{enumerate}
\item \textbf{Remote sensing inference}. SOC is a subsurface property that cannot be measured directly with remote sensing instruments. Remote sensing inference tools use correlations with underlying SOC levels by observing surface features (such as vegetation cover, soil texture, and moisture content).
\item \textbf{Scalability and efficiency}. Traditional in situ soil sampling for SOC measurements is time-consuming and costly, especially in larger regions. Remote sensing allows for large spatial data coverage, meaning it can detect changes in SOC (fundamental for its stock and dynamics).
\item \textbf{Real-time monitoring}. With remote sensing inference tools, it becomes possible to monitor SOC continuously or at regular intervals, allowing for (near) real-time data collection, without the need for continuous field sampling.
\item \textbf{Predictive modeling}. The platform applies statistical and ML models to predict SOC, using remotely sensed data to interpolate between regions lacking direct soil sampling or to extrapolate estimates from regions with little or no in-field SOC data.
\end{enumerate}

The rest of the paper is divided as follows. Section~\ref{sec:SOCPredRev} summarizes the literature on SOC prediction. Section~\ref{sec:SOCGenC} explains key concepts in SOC measurement. Section~\ref{sec:Methodology} outlines the strategy behind the proposed application. Section~\ref{sec:sample:dtcapabilities} describes the architecture and implementation details. Section~\ref{sec:Discussion} analyzes the application's core strengths and potential limitations, and, lastly, Section~\ref{sec:Conclusion} lays out the conclusions and future work.

%%%%%%%%%%%%%%%%%%%%%%%%%%%%%%%%%%%%%%%%%%%%%%%%%%%%%%%%%%%%%%%%%%%%%%%%%%%%%%%%%%%
\section{SOC prediction}
%%%%%%%%%%%%%%%%%%%%%%%%%%%%%%%%%%%%%%%%%%%%%%%%%%%%%%%%%%%%%%%%%%%%%%%%%%%%%%%%%%%
\label{sec:SOCPredRev}
WALGREEN uses spectral information obtained from the Landsat and Sentinel satellite missions thanks to its integration with Google Earth Engine (GEE) and the Copernicus Data Space Ecosystem via the Sentinel Hub. The Landsat program is a joint NASA and United States Geological Survey  program. The most recent Landsat satellite (Landsat 9) was launched on 27 September 2021. In Europe, the first Sentinel satellite (European Space Agency) was Sentinel 1, a polar-orbiting, all-weather, day-and-night radar imaging mission for land and ocean services. Sentinel-1A was launched on 3 April 2014. Another mission has been launched since 2014.

Research on SOC estimation has a long history (dating back even to~\cite{Waksman1926123}). To estimate SOC, onsite measurements at various geographic points and satellite image observations are used to train ML systems for SOC prediction. Thus, costly onsite measurements that require extensive field work can be minimized.

Public frameworks for remote sensing are currently available, including \href{https://github.com/moienr/SoilNet/blob/main/README.md}{SoilNet}, a hybrid transformer-based framework with self-supervised learning for large-scale SOC prediction, and CataEx~\cite{domej2025}, a multi-task export tool for the GEE data catalog. SoilNet lacks a public interface, while CataEx only covers  the GEE side and supports multiple small- and large-scale SOC mapping efforts in different regions. Also, using these frameworks requires knowledge of Python, R, and other programming languages.

The research on SOC estimation, with the use of remote sensing techniques~\cite{Angelopoulou2019, van2023remote,Radocaj2024,Falahatkar2014507}, shows how using remote sensing data to estimate SOC regionally or nationally offers a solution to the COP21 challenge~\cite{Ladoni201082,neofytou2024review,Zizala2019}, especially if we look at the improvements in hardware that have increased the speed of high-performance computing methods~\cite{wang2024distributed}.

Sensors mounted on satellites let us use remote sensing techniques in the visible-near infrared–shortwave infrared (400–2500 nm) region. This approach offers a more direct, cost-effective, and rapid way to estimate key indicators for soil monitoring~\cite{angelopoulou2019remote} .

Numerous studies have applied SOC prediction to specific areas (Rodríguez Martín~\cite{RodriguezMartin2019}, Zhang \cite{zhang2017prediction,zhang2019prediction}, Ferreira \cite{ferreira2023predicting}  and Mondal \cite{mondal2017spatial}), including comparisons of remote sensing devices. Maxwell \cite{maxwell2023global} collated a global dataset of tidal marsh soil organic carbon (MarSOC) from 99 studies, providing the tidal MarSOC database with 17454 datapoints. Wang~\cite{wang2018104} aimed to improve the potential of Analytical Spectral Devices hyperspectral data and Landsat Operational Land Imager data for predicting soil organic matter content in the bare topsoil of the Ebinur Lake Wetland National Nature Reserve in northwest China. Francos~\cite{francos2024mapping} used hyperspectral remote sensing to model the SOC stock in the Sele River plain located in the Campania region in southern Italy.

WALGREEN offers basic ML methods implemented as software as a service (SaaS) for prediction models. Numerous studies have been conducting in this area. For instance, Neofytou~\cite{neofytou2024review} reviewed ML and deep learning methods used in the last five years, analyzing  $52$ papers and their metric ($R^{2}$) ranges. New two-stage pipelines for remote SOC estimation have also been proposed~\cite{Pavlovic2024}. Liu~\cite{Liu2024} compares and analyzes the performance of partial least squares regression, support vector regression (SVR), random forest (RF), and Gaussian process regression (GPR) for SOC inference.

The first version of our application, presented at~\cite{aroca2024walgreen}, only had Copernicus Sentinel Hub integration, and map handling was different to in the current version. In addition, ML algorithms have now been integrated following a SaaS model.

%%%%%%%%%%%%%%%%%%%%%%%%%%%%%%%%%%%%%%%%%%%%%%%%%%%%%%%%%%%%%%%%%%%%%%%%%%%%%%%%%%%
\section{SOC general concepts}
\begin{figure}
    \centering
    \includegraphics[width=1\linewidth]{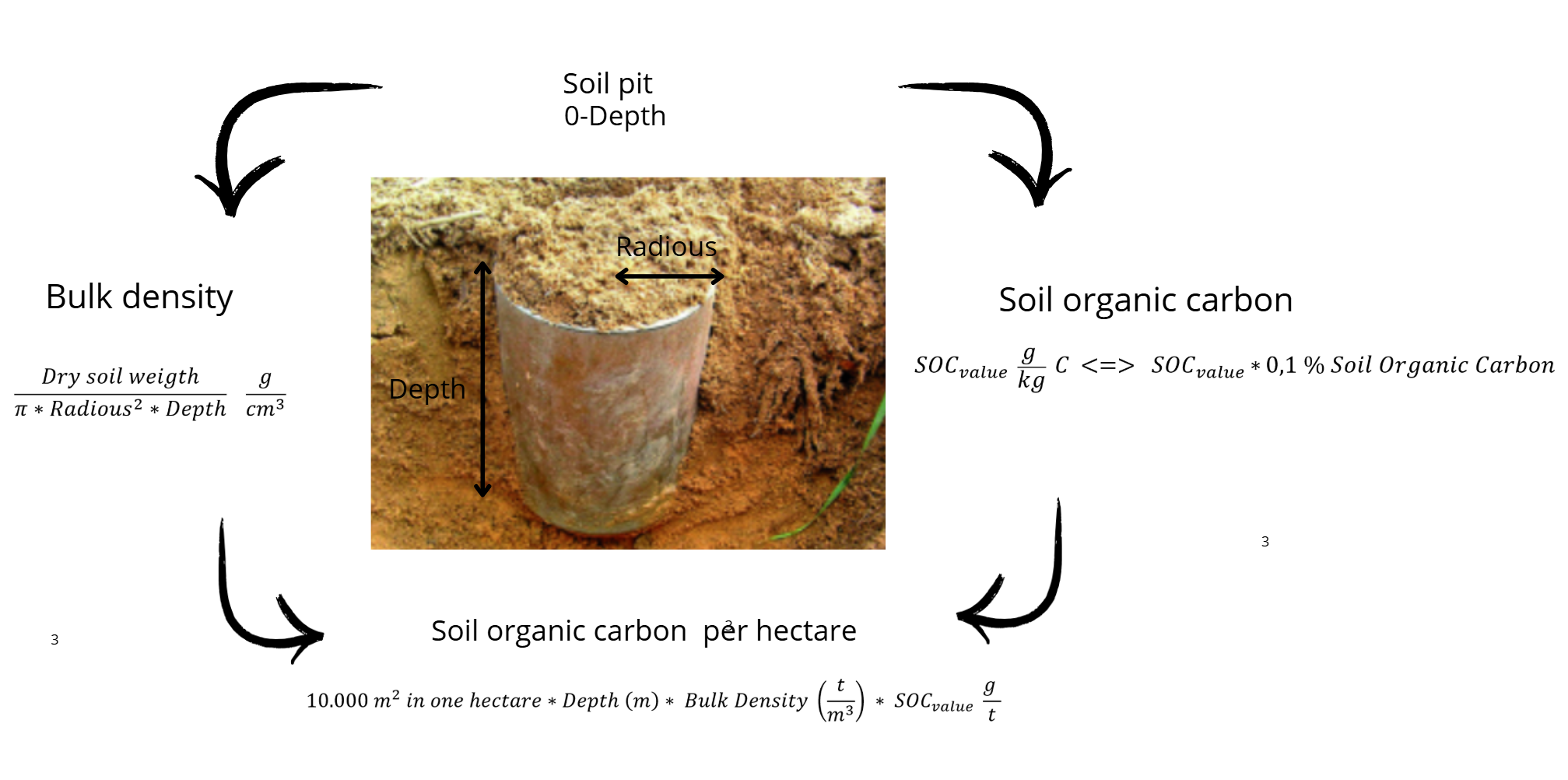}
    \caption{SOC measurement description}
    \label{fig:SOC measurement}
\end{figure}

%%%%%%%%%%%%%%%%%%%%%%%%%%%%%%%%%%%%%%%%%%%%%%%%%%%%%%%%%%%%%%%%%%%%%%%%%%%%%%%%%%%
\label{sec:SOCGenC}
While the aim of this paper is to present a software application and explain its main features and functionality, a basic understanding of SOC processes, methods, and practices can help clarify key concepts and enhance the effective use of the application. The core components of the SOC estimation process are:
\begin{enumerate}{}{}
\item \textbf{Measurement and monitoring}. This includes methods for collecting soil samples at different depths and locations, laboratory analysis, and remote sensing and modeling.
\item  \textbf{Drivers and dynamics}. The SOC value and (spatiotemporal) distribution are affected by different drivers, including organic matter proxies (crop residues, roots, and manure) and natural drivers like decomposition, respiration, erosion processes, and climate factors.
\item  \textbf{SOC indicators}. These include bulk density, soil texture, and organic matter content. Bulk density is the dry weight of a known soil volume; soil tests for organic carbon normally report a \% total SOC. Using a measure of bulk density, the amount of carbon per hectare in a given depth of soil can be calculated as shown in Figure~\ref{fig:SOC measurement}, WALGREEN uses $g/cm^3$ as the SOC measurement standard unit.
\end{enumerate}
Changes in SOC typically occur over many years and are difficult to detect due to their relatively small magnitude compared to the total carbon stock in the soil. For example, soils in Western Australia contain between 20 and 160 tonnes of carbon per hectare. A typical wheat crop that produces 2 tonnes per hectare can retain between 0.1 and 0.5 tonnes of organic matter per hectare annually, which is less than 1\% of the total SOC stock. Therefore, it would take more than ten years to detect a significant change in SOC in typical grain cropping systems.

The capacity of soils to store organic carbon represents a key function of soils \cite{Wiesmeier2019149}. Conventional SOC measurements are obtained through direct soil sampling followed by laboratory analysis, which is laborious and costly. Because of the potentially high spatial heterogeneity of SOC, direct sampling is laborious, expensive, and time-consuming. However,  remote sensing tools are powerful for providing SOC estimates over large spatial extents at a considerably lower cost with less labor investment.

%%%%%%%%%%%%%%%%%%%%%%%%%%%%%%%%%%%%%%%%%%%%%%%%%%%%%%%%%%%%%%%%%%%%%%%%%%%%%%%%%%%
\section{Methodology}
%%%%%%%%%%%%%%%%%%%%%%%%%%%%%%%%%%%%%%%%%%%%%%%%%%%%%%%%%%%%%%%%%%%%%%%%%%%%%%%%%%%
\label{sec:Methodology}
WALGREEN is built on the Model-View-Controller (MVC) architecture and provides an interface that enables users to leverage remote sensing technologies without requiring extensive technical expertise. With it, users can interact with different Earth observation platforms to access satellite GeoTIFFs and reflectance datasets, manage SOC public datasets, manage SOC user datasets, and predict SOC.

The MVC development adheres to the separation of concerns principle, a design model that divides an application into different units with minimal functionality overlap. The following are the key components of the MVC architecture as applied in our development \cite{necula2024exploring}:
\begin{enumerate}{}{}
\item \textbf{Model}: manages the application's data, generally using JPA (Java Persistence API) with Hibernate ORM (object-relation mapping). Hibernate is primarily used for mapping Java objects to database tables.
\item \textbf{View}: implements MVC in Spring Boot with Thymeleaf, rendering views as HTML templates. Thymeleaf enables seamless injection of dynamic content into HTML while maintaining compatibility with JavaScript.
\item \textbf{Controller}: handles user interactions, processes requests, interacts with the model, and updates the view accordingly.
\end{enumerate}
The WALGREEN web app architecture is based on a platform with a front-end, a back-end, and a RESTful API.  The RESTful API functionality entails ML and the integration of the Earth observation platform with the Copernicus Sentinel Hub API and the GEE API (see~\ref{sec:sample:cshvsgee} for more details).

Integrated Earth observation platforms are powerful tools for working with Earth observation data and geospatial analysis. However, even though these platforms share some characteristics at the functional level, they cater to slightly different needs and have unique features. What WALGREEN aims to do is build a similar navigation map for downloading GeoTIFF files and accessing data in Sentinel Hub and GEE.

\begin{figure}[h]
    \centering
    \includegraphics[width=1\linewidth]{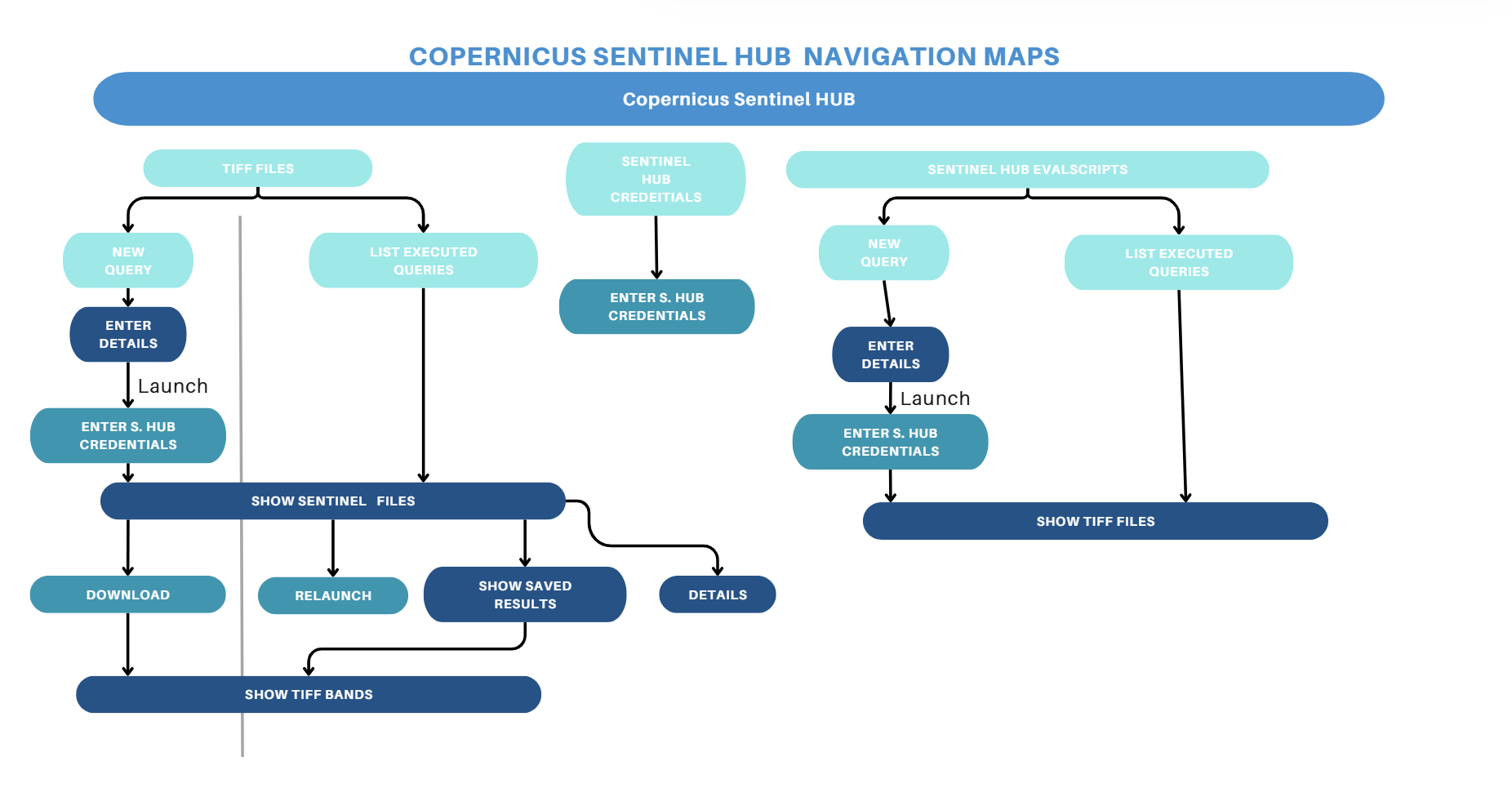}
    \caption{Copernicus Sentinel Hub navigation map}
    \label{fig:CSH_nav_map}
\end{figure}

\begin{figure}[h]
    \centering
    \includegraphics[width=1\linewidth]{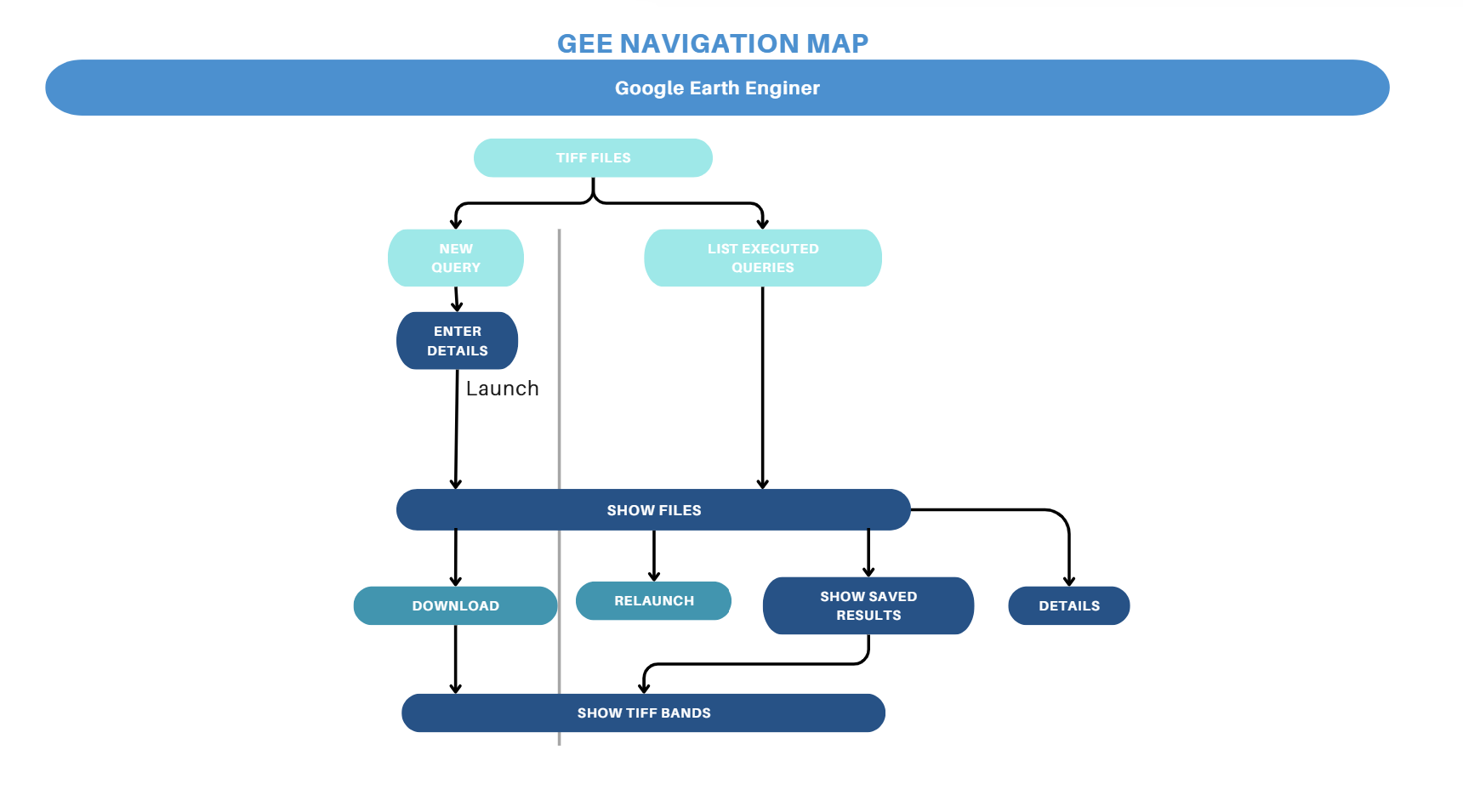}
    \caption{GEE navigation map}
    \label{fig:GEE_nav_map}
\end{figure}

Copernicus Sentinel Hub API and the GEE API have different integration APIs and workflows in terms of design and programming integration. For example, in the navigation maps, one difference is found when handling credentials for calling APIs. Copernicus Sentinel Hub API (see Figure~\ref{fig:CSH_nav_map}) requires authentication, and while for researchers access is free, it has low \href{https://documentation.dataspace.copernicus.eu/Quotas.html} {quotas and other limitations}. In contrast, the GEE API only requires unique project credentials integrated at the application level. While it does impose some \href{https://developers.google.com/earth-engine/guides/usage}{rate limits}, by taking care of the main one, i.e., the number of requests per minute, it is possible to not have to prompt users for credentials (see Figure \ref{fig:GEE_nav_map}). The integration details are outlined in Section \ref{sec:sample:dtcapabilities}.

\begin{figure}[h!]
    \centering
    \includegraphics[width=1\linewidth]{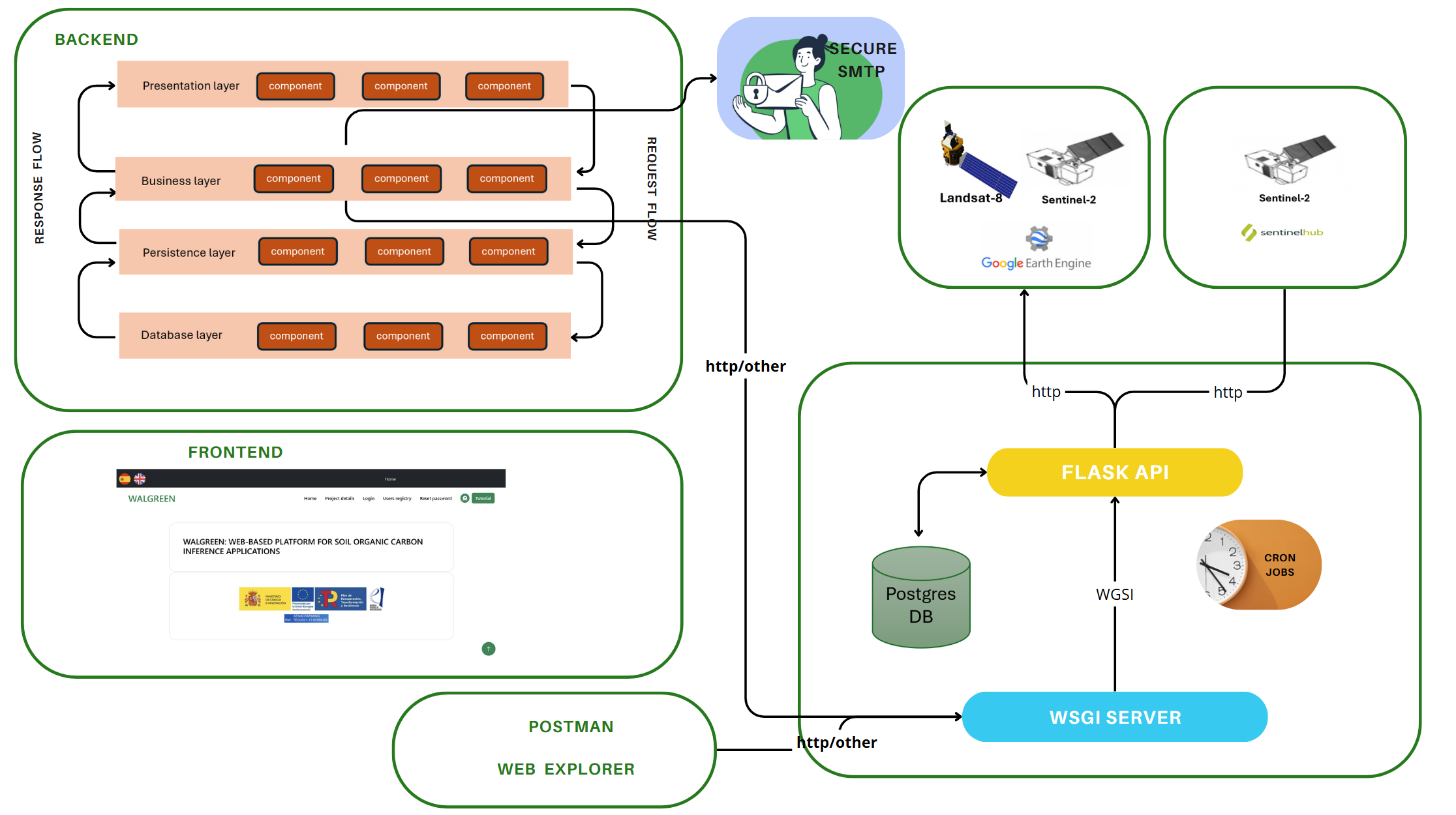}
    \caption{WALGREEN architecture}
    \label{fig:architecture}
\end{figure}
\begin{figure}[h]
    \centering
    \includegraphics[width=1\linewidth]{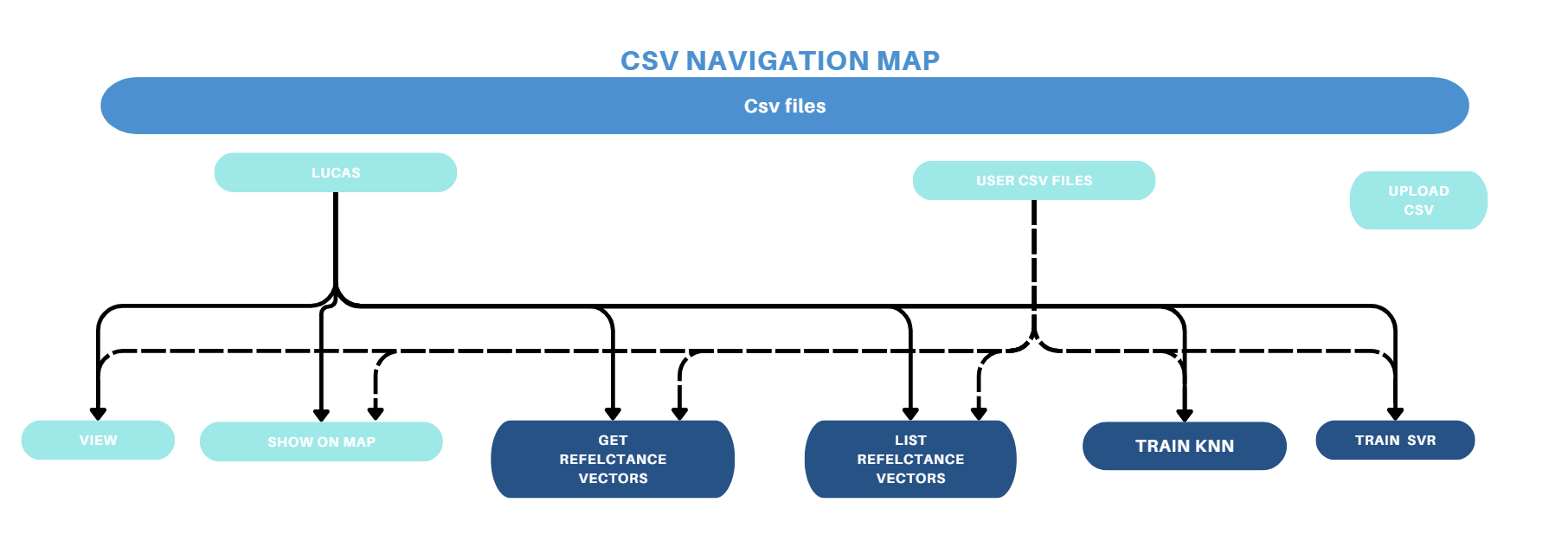}
    \caption{CSV user files navigation map}
    \label{fig:csv_nav_map}
\end{figure}


%%%%%%%%%%%%%%%%%%%%%%%%%%%%%%%%%%%%%%%%%%%%%%%%%%%%%%%%%%%%%%%%%%%%%%%%%%%%%%%%%%%
\section{ Design and tool capabilities}
%%%%%%%%%%%%%%%%%%%%%%%%%%%%%%%%%%%%%%%%%%%%%%%%%%%%%%%%%%%%%%%%%%%%%%%%%%%%%%%%%%%
\label{sec:sample:dtcapabilities}
WALGREEN offers several features, including CSV file handling (Figure \ref{fig:CSH_nav_map}) and  integration with Copernicus Sentinel Hub (Figure \ref{fig:copernicusshwf}) and GEE (Figure \ref{fig:GEE_nav_map}). It also provides a RESTful API  integrated into the main application that offers functionality such as:
\begin{enumerate}{}{}
\item \textbf{GEE integration}.
The GEE public data catalog includes a variety of standard Earth science raster datasets. The GEE API provides radar images from the Copernicus program, including rain- and snow-free radar images from Sentinel-1A and 1B, and high-resolution optical images from Sentinel-2A and 2B. Zhou ~\cite{Zhou2023}, Coskun \cite{Coskun2024499}, and Javidan~\cite{javidan2024soil} represent three examples of GEE integration.
Users can access a wide range of satellite and environmental data. The platform simplifies geospatial analysis by automatically handling data projection, scaling, and composition based on user-specified parameters.
\item \textbf{Copernicus integration}.
The Copernicus Data Space Ecosystem offers multiple APIs, ranging from those for catalog access, product downloads, and visualization to processing  web services such as  the SpatioTemporal Access Catalog, openEO, and Sentinel Hub APIs.
WALGREEN is integrated with Sentinel Hub APIs and allows users to access raw satellite data, rendered images, statistical analysis, and other features.
\item \textbf{Data science and ML}.
WALGREEN has ML methods implemented as a SaaS. RFs, SVR, and k-NN are trained using the Land Use/Cover Area Frame Survey (LUCAS) datasets or any potential user datasets. It allows users to obtain predictions of SOC values or SOC-predicted values for specific areas using the existing, previously trained algorithms.

\end{enumerate}

Although WALGREEN lets users work with its  datasets, one of the aims is to use public datasets to train predictive models that non-specialists can use for SOC prediction. For this, the LUCAS topsoil database is used~\cite{Toth20137409,Wiesmeier2019149}. The LUCAS spectral data resampled according to the multispectral bands are used to infer SOC (van Wesemael \cite{van2024european}, Castaldi \cite{Castaldi2019,Castaldi2018592}, Kakhani \cite{Kakhani2024}, Petito \cite{Petito2024}).

\subsection{Copernicus Sentinel Hub vs GEE}
\label{sec:sample:cshvsgee}

Given WALGREEN's functionality requirements, the integration process has helped us understand the strengths and weaknesses of each platform from different points of view, as summarized in Table~\ref{table1}. Both tools complement each other. The choice between them depends on the project's requirements and the user's technical expertise.

\begin{longtable}{ |p{2cm}||p{6cm}|p{6cm}|  }
\caption{Sentinel Hub API vs GEE API.\label{table1}}\\
 \hline
Feature&Sentinel Hub API&GEE API \\
  \hline\hline
\cellcolor[gray]{0.8}&{\vrule width 1pt}\cellcolor[gray]{0.9}\xmark Mainly used when efficient access and streaming of satellite imagery is required.&\cellcolor[gray]{0.9}\cmark Mainly used for geospatial analysis and research purposes. \\
\cellcolor[gray]{0.8}&\xmark Commercial and enterprise use with paid plans; has paid, low-cost limit licenses.&\cmark Free license for research and non-commercial use; requires approval for commercial distribution. \\
\multirow{-3}{4em}{\cellcolor[gray]{0.8}Synopsis}&\cellcolor[gray]{0.9}\cmark Streaming-based, optimized for fast imagery access, also provides cloud-based analytics with JavaScript/Python APIs. &\cellcolor[gray]{0.9}\cmark Cloud-based analytics with JavaScript/Python APIs. \\

&\cmark Sentinel, Landsat, commercial data options.&\cmark Access to satellite imagery (Sentinel, Landsat, MODIS, commercial). \\
&\cellcolor[gray]{0.9}\cmark Cloud-based processing without downloading data locally to the browser.&\cellcolor[gray]{0.9}\cmark Cloud-based processing, extensive geospatial catalog, including satellite imagery, climate, terrain, and socio-economic data.\\
\multirow{-3}{4em}{Data access}&\xmark Requires downloading the TIFF file to get reflectance values.&\cmark TIFF file analysis within the API server lets the user download historical dataset reflectance values without downloading files.  \\

\cellcolor[gray]{0.8}&\cellcolor[gray]{0.9}\cmark Optimized for high-speed delivery via WMS/WMTS and API endpoints.&\cellcolor[gray]{0.9}\cmark On-the-fly image processing (e.g., RGB composites, NDVI calculations). \\
\cellcolor[gray]{0.8}&\xmark Supports batch processing; focuses on real-time analysis but less suited for large-scale analysis.&\cmark  Comprehensive geospatial computing environment for large-scale analysis. \\
\multirow{-3}{4em}{\cellcolor[gray]{0.8}Processing}&\cellcolor[gray]{0.9}\xmark Cloud-based analytics with JavaScript/Python APIs with optimized capabilities for fast imagery access.&\cellcolor[gray]{0.9}\cmark Cloud-based analytics with JavaScript/Python APIs with optimized capabilities for batch image processing. \\

&\xmark When focusing on satellite imagery, requires runtime generation of custom scripts (eval scripts) for querying.&\cmark Straightforward for users focused on satellite imagery. Functionality included in API. \\
\multirow{-2}{4em}{Usability}&\cellcolor[gray]{0.9}\cmark Accessible through Python, OGC APIs, EO Browser, and a valuable custom scripting language with high processing capabilities.&\cellcolor[gray]{0.9}\cmark Accessible through JavaScript (Code Editor),  Python and browser, also supports custom scripts and ML.

\\
\hline
\end{longtable}

%================================================================================
\subsection{Architecture}
%================================================================================
\label{sec:sample:architecture}
The WALGREEN architecture (Figure~\ref{fig:architecture}) is based on an MVC front-end and back-end coded in Java using Spring Boot, Hibernate, and Spring Data JPA with Thymeleaf. The Earth observation side is based on an API coded in Python using Web Server Gateway Interface (WSGI) and Flask.

To provide an in-depth exploration of Spring Boot---a highly sought-after framework in modern web development---Hubli~\cite{hubli2023efficient} gave a thorough overview of its core principles, key features, and best practices.

Flask is a popular Python micro-framework for developing lightweight applications and RESTful APIs. It provides only the essential components, allowing developers to integrate additional tools as needed. In the WALGREEN application, a Flask API has been deployed using WSGI and an Apache server. The uWSGI version serves as the master daemon, managing and supervising Python worker processes. It handles time-capped requests, worker recycling, background tasks, cron jobs, timers, logging, automatic reloads on code changes, and run-as privileges.

Depending on the different features, API methods can invoke Copernicus Sentinel Hub or GEE APIs to set up synchronous or asynchronous interactions to serve JSON format data structures when called from the back-end app side.
\begin{figure}
    \centering
    \includegraphics[width=1\linewidth]{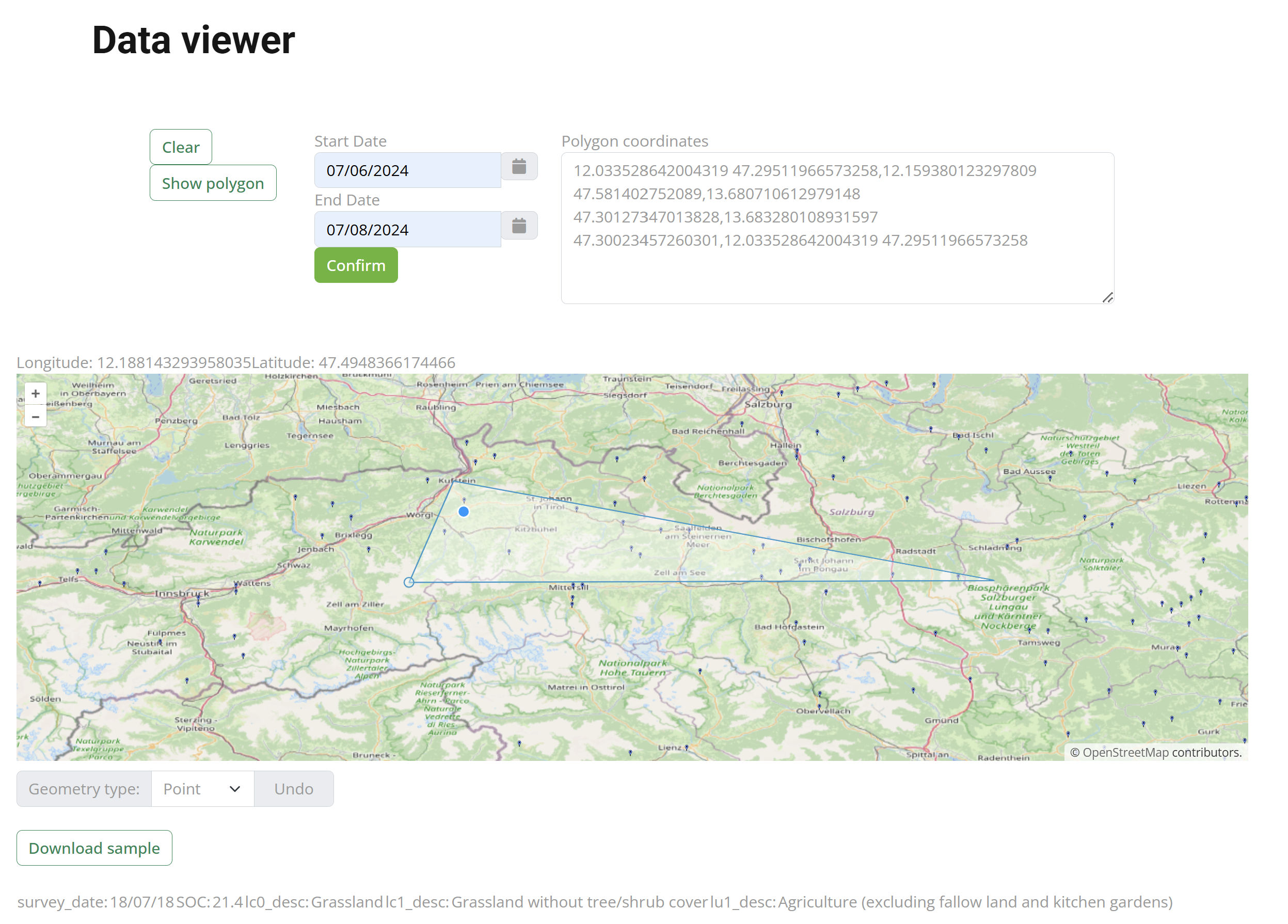}
    \caption{The SOC viewer displays every point in the CSV file and offers information on each point}
    \label{fig:mapvierer}
\end{figure}

%================================================================================
\subsection{CSV file processing}
\begin{figure}
    \centering
    \includegraphics[width=1\linewidth]{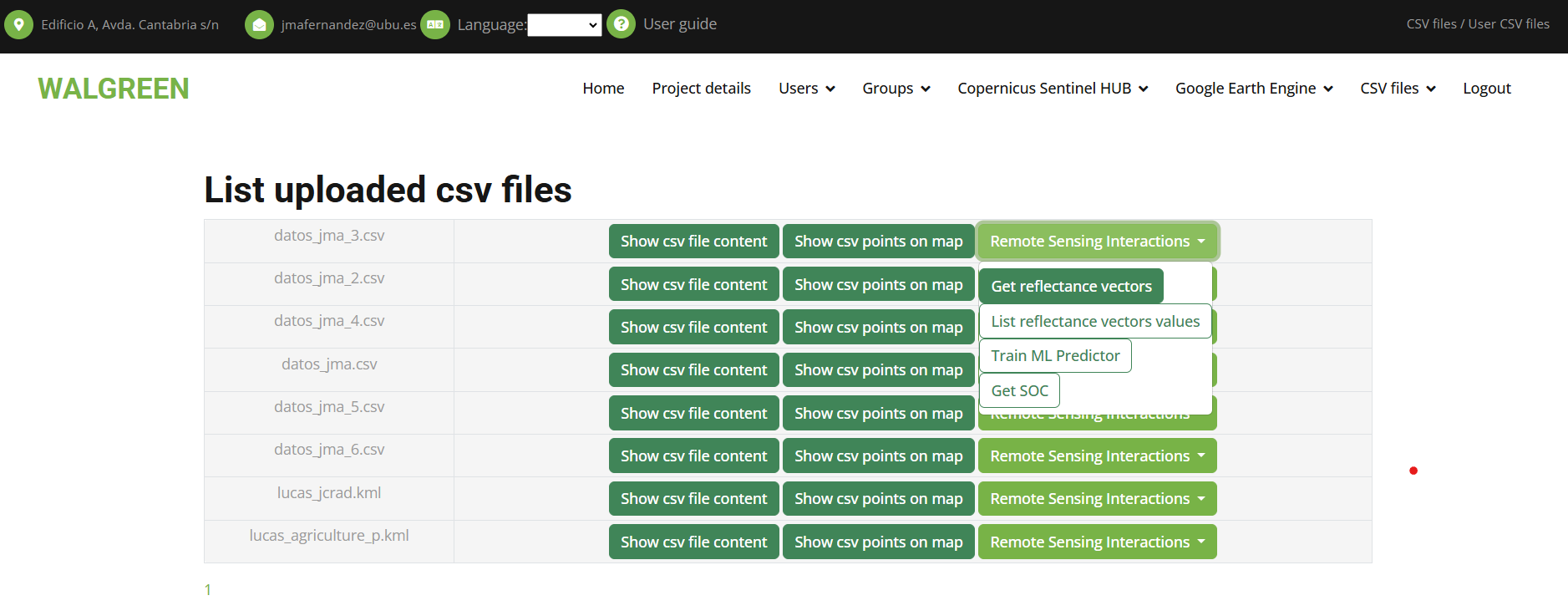}
    \caption{CSV files options when training ML methods}
    \label{fig:csv options 1}
\end{figure}

\begin{figure}
    \centering
    \includegraphics[width=1\linewidth]{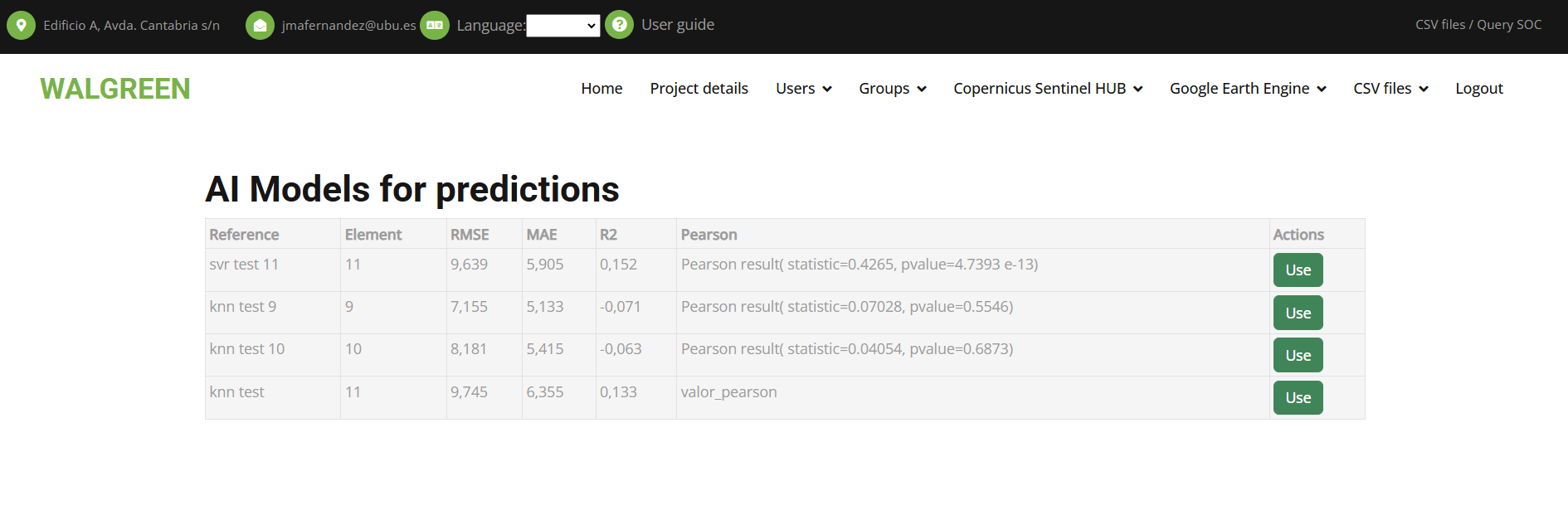}
    \caption{ML predictor options list}
    \label{fig:predictors}
\end{figure}

%================================================================================
\label{sec:sample:desing_csv}
CSV files (Figure~\ref{fig:csv_nav_map}) allow users to interact with the web app, use the tool, and take advantage of the functionality provided. To improve navigation, there is just one screen for uploading the files. Once a CSV file is uploaded, its content is saved to the database, and depending on the processing stage, different actions might be taken (as shown in Figure~\ref{fig:csv options 1}). Each option will lead to other screens so that the user can launch the following actions:
\begin{itemize}
\item \textbf{Download and query public and shared datasets}.
WALGREEN lets users download and view the datasets used to train the public ML inference methods and the users' datasets.
\item \textbf{Display SOC values on a map}. WALGREEN has its own integrated visor for displaying the points in a CSV file (as shown in Figure~\ref{fig:mapvierer}), allowing users to zoom in and out, rotate, and apply other functions. Section~\ref{sec:openlayer} provides technical references for this functionality. The Open Street Map version for Node.js has been modified to work with JavaScript, Thymeleaf, and Java.

\item \textbf{Query SOC values}. Users can choose from a list an ML predictor and use it to query SOC for a set of coordinates or an entire CSV file.
\item \textbf{Train ML methods}:  Users can train ML models to improve the predictions given by WALGREEN (Figure~\ref{fig:predictors}). Each predictor receives a set of metrics so users can check the accuracy of each.

The metrics used are root-mean-squared error (RMSE), mean absolute error (MAE), coefficient of determination (also known as R-squared or $R^2$), and the Pearson correlation index. RMSE is optimal for normal (Gaussian) errors; MAE is optimal for Laplacian errors. When errors deviate from these distributions, other metrics are better \cite{hodson2022root}. $R^2$ should be used as the standard metric to assess regression quality in any scientific domain \cite{chicco2021coefficient}.

\end{itemize}

%================================================================================
\subsection{Satellite data and TIFF image integration}
%================================================================================
Combining satellite data with TIFF (tagged image file format) images is a common technique in geospatial analysis since TIFF files, particularly GeoTIFFs, are among the most common formats for storing raster geospatial data. Satellite data includes multispectral, hyperspectral, and radar imagery, as well as reflectance, temperature, and elevation, which are just a few examples of the values a single pixel can represent. These images are stored as raster data.

TIFF is a flexible image format, while GeoTIFF is an extension that embeds geospatial metadata (e.g., coordinate reference systems and map projections) in a TIFF file.
The workflow used to integrate satellite data and TIFF images has three steps: (1) data acquisition; (2) pre-processing; (3) visualization and analysis.
Depending on the data provider (each with different APIs, strengths, and limitations), each step referred to above has different internal integration workflows, which are described in Sections~\ref{sec:copernicus} and~\ref{sec:gee}.\\

%================================================================================
\subsection{Copernicus: Satellite data and TIFF image integration}
%================================================================================
\label{sec:copernicus}
WALGREEN integration with the Copernicus Sentinel Hub provides three functions: (1) querying individual TIFF files, (2) launching eval scripts, and (3) querying multiple TIFF files for a CSV file that contains SOC values. No action can be taken if the user is not correctly authenticated using access token credentials from the Copernicus platform. Once the credentials have been verified, the user can browse the menu options.

The Copernicus Sentinel-2 mission consists of two polar-orbiting satellites in the same sun-synchronous orbit with a $180^{\circ}$ phase difference. This is vital for understanding the TIFF image download processes.

The first two functions let users download single images. The first lets users download one TIFF file per band, while the second lets users get a single TIFF file using an eval script (or ``custom script''). This script consists of JavaScript code that defines how Sentinel Hub must process the satellite data and what values the service is to return. The API enforces the image width to be an integer between $1$ and $2500$ when launching a request.

\begin{figure}
    \centering
    \includegraphics[width=0.9\linewidth]{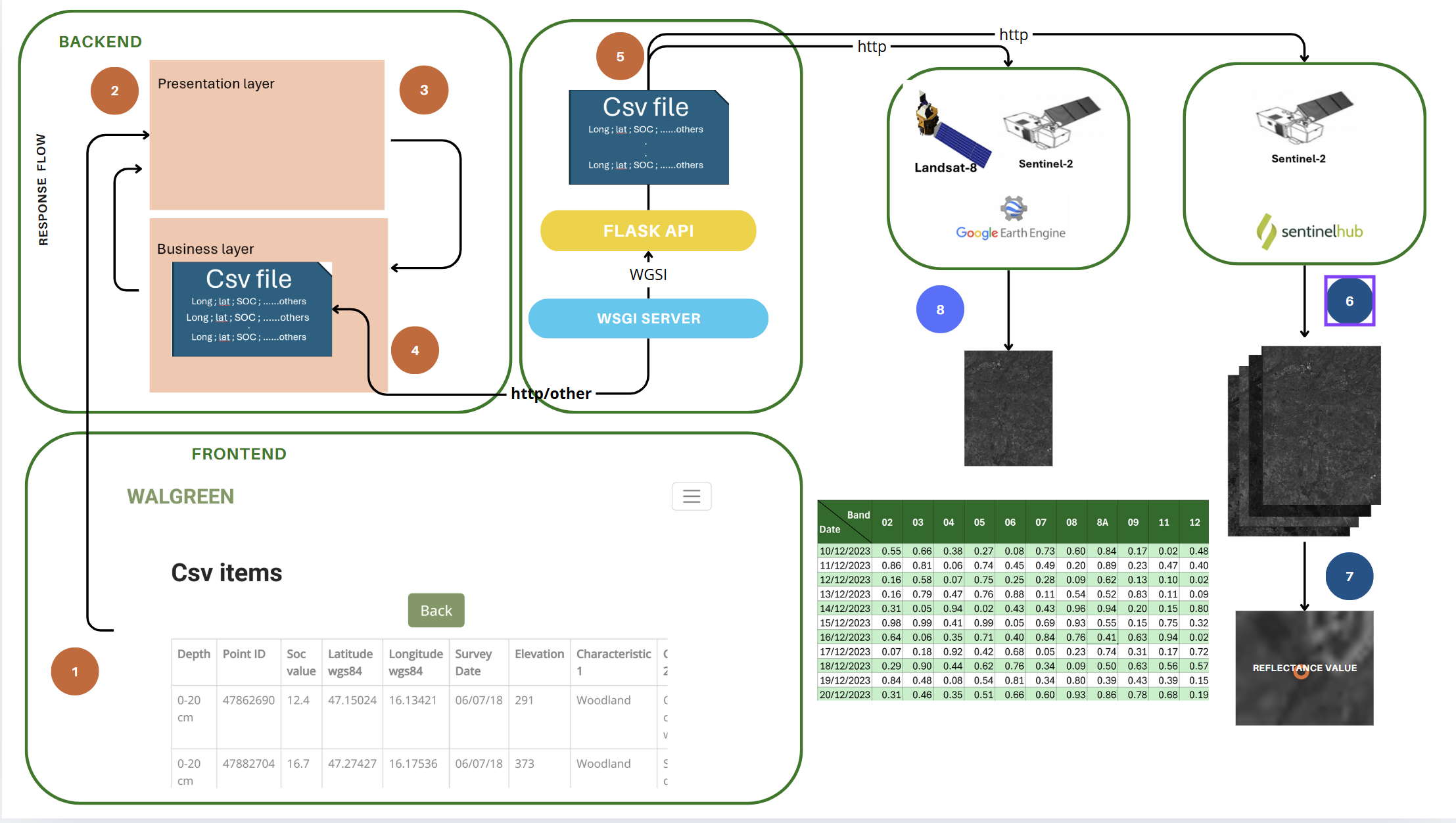}
    \caption{Data and image integration: Common steps: 1. Select CSV. 2 and 3. Submit action to get reflectance. 4. Send CSV file to Python sever to get reflectance (async call). 5. For each line, get TIFF images for each band. Copernicus Sentinel Hub steps: 6. Save TIFF images. 7. Get reflectance vector. GEE steps: 8. Save TIFF images and reflectance vector matrix}
    \label{fig:copernicusshwf}
\end{figure}
The most notable functionality is the third because it entails not only data acquisition but also TIFF data pre-processing (see Figure~\ref{fig:copernicusshwf}), with each TIFF file being processed to get a reflectance value for the coordinates. The final result obtained is a matrix saved to the database with a reflectance vector (one reflectance per satellite band) for each point in the CSV file.

The reflectance value range can be handled in different ranges, such as INT8---signed 8-bit integers (values range from -128 to 127); UINT8---unsigned 8-bit integer (values range from 0 to 255); INT16---signed 16-bit integer (values range from -32768 to 32767); UINT16---unsigned 16-bit integer (values range from 0 to 65535); FLOAT32---32-bit floating point (values have effectively no limits); and AUTO (default)---values range from 0-1, which will then automatically be stretched from the interval [0, 1] to [0, 255] and saved as a UINT8 raster.

WALGREEN can create an eval script (in real-time) depending on the band and the unit the user selects for downloading TIFF images. We recommend using FLOAT32 for processing images and AUTO for TIFF files to be displayed in an image viewer. The final result is a dataset consisting of geographic coordinates (longitude, latitude), an SOC value, and the corresponding reflectance vector for each point that can be downloaded to train external ML methods or ML methods offered as SaaS.\\

%================================================================================
\subsection{GEE: Satellite data and TIFF image integration}
%================================================================================
\label{sec:gee}
WALGREEN integration with GEE lets users query multiple TIFF files for a CSV file containing SOC values. WALGREEN does not need the user's credentials for GEE. The system can call the GEE platform by adhering to its rate limits on the number of calls per minute.

The integration process for this functionality (see Figure~\ref{fig:copernicusshwf}) starts by a CSV file being read. For each line (SOC data point) and Sentinel-2 (after the application of a cloud mask), WALGREEN asks for the list of TIFF images with harmonized data. Without downloading any TIFF image, the tool gets the reflectance value for the specified point coordinates, obtaining a matrix that is saved to the database as the final result, containing a reflectance vector.

\begin{figure}
    \centering
    \includegraphics[width=1\linewidth]{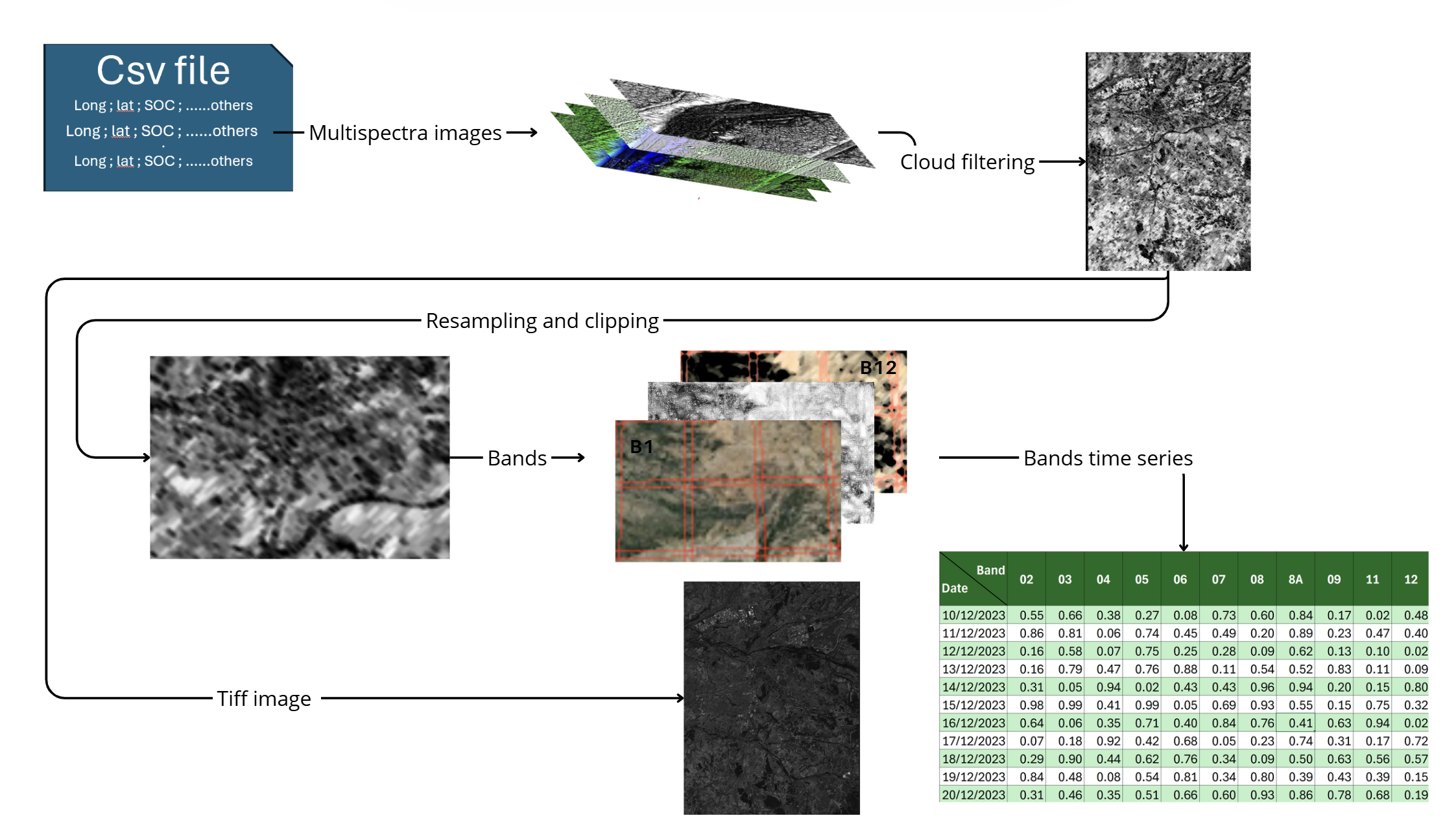}
    \caption{GEE cloud filtering, scaling, resampling, and clipping }
    \label{fig:geel}
\end{figure}
As suggested in recent articles~\cite{Valerio2024}, the scaling, resampling, and clipping are performed for each TIFF file the via the GEE API (see Figure \ref{fig:geel}). The system downloads only a unique TIFF file with the best cloud masking value.

%================================================================================
\subsection{OpenLayers, JavaScript and Thymeleaf integration}
%================================================================================
\label{sec:openlayer}
WALGREEN uses OpenLayers (version 10.2) to integrate map management. Initially, it worked with JavaScript simple maps, but further functionality was needed, for instance, support for Keyhole Markup Language (KML) files, drawing points and polygons, zooming in and out, etc. This library was developed to be used with Node.js, not JavaScript and Thymeleaf. We also modified the required OpenLayers libraries to make them work without Node.js, draw polygons and points, drop KML files, download and view TIFF files, etc.

We did this by installing a spare ``Node Modules'' directory using the libraries, copying the js files to the static folder in the Spring Boot structure, and integrating files by modifying import statements or using ``importmap'', an ECMAScript feature that facilitates specifying the URL from which modules are to be downloaded, using mechanisms such as \textit{import} or \texttt{import()}, where a bare \textit{import} is used.

%================================================================================
\subsection{Linear interpolation and filtering}
%================================================================================
\label{sec:Kalman}
As mentioned in Sections \ref{sec:sample:dtcapabilities} and \ref{sec:gee}, WALGREEN allows users to handle public or private datasets with SOC measurements for a particular location on a specific date. As stated in Section \ref{sec:sample:desing_csv}, a reflectance vector per location is necessary to train regression (or other) methods; it is not always possible to get reflectance vectors for the specified date. In these cases, it might be necessary to infer it.

Kalman filter and linear interpolation are both methods used to estimate or predict values in a sequence, but they serve different purposes and operate under different assumptions. The first is recommended for assessing the state of a dynamic system over time, often in the presence of noise. In contrast, linear interpolation is recommended for estimating unknown values between two known dates in a static dataset. Other advantages of the Kalman filter include the ability to handle noisy observations and there being no need to provide labeled training data.

Given a known spatial point and date, a WALGREEN query for GeoTIFFs in a date range around the target date is not guaranteed to get the reflectance vector. In this case, we will not always look for unknown values between two known dates. The query date is often earlier or later than the lower or upper dates for the GeoTIFFs found. This is why the Kalman filter might be used for an observation in one dimension (the reflectance value in a particular band).

In our case, the Kalman filter has been implemented using the \texttt{pykalman} library, coding the required use of the \texttt{KalmanFilter} method by specifying both an initial value from which the filter can be derived and the size of observation space. The technical solution also integrates the Kalman Smoother and the EM algorithm (model parameter inference).

Kalman Smoothers are used to estimate the state of a linear dynamical system from noisy measurements. It incorporates “future” measurements as well as past ones. On the other hand, EM is an iterative algorithm for computing maximum likelihood estimates or posterior modes in problems with incomplete data. In its general form, the E (expectation) step computes the conditional expectation of the complete-data log-likelihood function given the observed data and the parameter estimate from the previous iteration. The M (maximization) step maximizes this expected log-likelihood function to determine the next parameter estimation stage \cite{meng2004algorithm}.

\begin{figure}[ht!]
    \centering
    \includegraphics[width=1\linewidth]{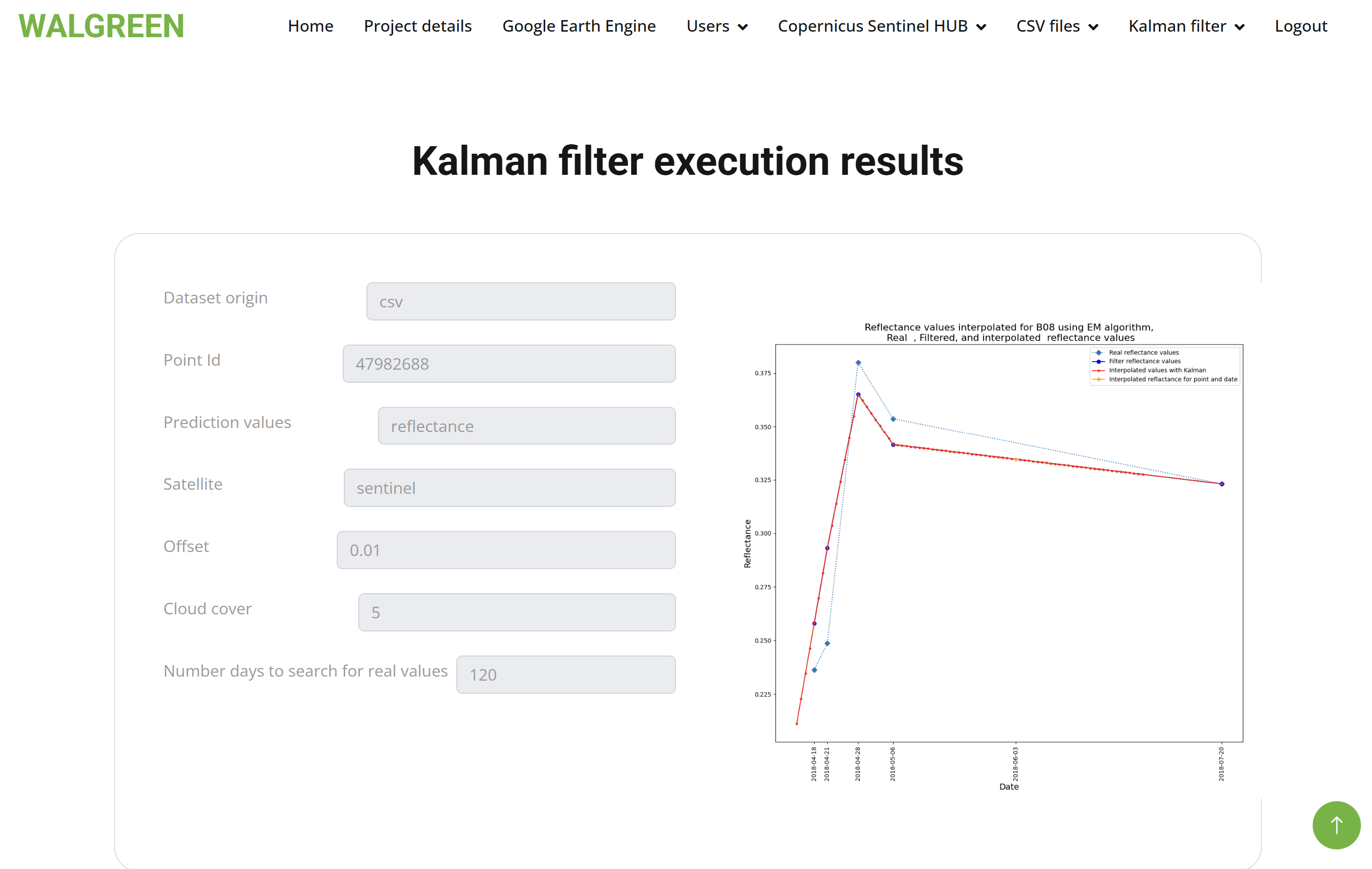}
    \caption{Interpolated band $08$ reflectance values obtained with the EM algorithm}
    \label{fig:kalman}
\end{figure}

The EM algorithm usually gives better results than Kalman Smoothers. In Figure \ref{fig:kalman}, given a known point and date, the actual, filtered, and interpolated reflectance values for band $08$ are shown. The reflectance interpolated value for that date is shown in orange.

%%%%%%%%%%%%%%%%%%%%%%%%%%%%%%%%%%%%%%%%%%%%%%%%%%%%%%%%%%%%%%%%%%%%%%%%%%%%
\section{Discussion}
%%%%%%%%%%%%%%%%%%%%%%%%%%%%%%%%%%%%%%%%%%%%%%%%%%%%%%%%%%%%%%%%%%%%%%%%%%%%
\label{sec:Discussion}
%Summarize the key findings in clear and concise language. ...
%Acknowledge when a hypothesis may be incorrect. ...
%Place your study within the context of previous studies. ...
%Discuss potential future research. ...
%Provide the reader with a “take-away” statement to end the manuscript.
Our platform gives users access to location and real-time SOC data point analysis, allowing different types of users (researchers, policymakers, and even farmers) to analyze soil health and carbon storage. This can help in sustainable agriculture practices, land restoration, and climate adaptation strategies.

The platform's main strength is its integrated functionality. Unlike other, previous platforms, which often focus on isolated functionality, i.e., only the remote sensing side for obtaining TIFF images and reflectance vectors, launching specific studies with ML methods on small areas, or creating only the framework side without a front-end, WALGREEN offers end-to-end functionality that lets users obtain SOC values without any previous technical expertise.

Despite its strengths, the framework has several limitations that must be addressed. These include overcoming the account limits imposed by remote sensing data providers, integrating additional ML algorithms, and, most notably, securing financial support for covering hardware expenses. Depending on the number of users and the ML algorithms in use, greater capacity might also be required for processing and data storage.

The platform has the potential to be widely adopted for querying and storing TIFF files and related datasets. It would allow users to relaunch and reproduce queries and experiments without consuming API resources more than once for the same task. Additionally, if information in Copernicus Sentinel Hub or GEE is altered or no longer available, the web app would retain records of downloaded TIFF files, datasets, and results.

%%%%%%%%%%%%%%%%%%%%%%%%%%%%%%%%%%%%%%%%%%%%%%%%%%%%%%%%%%%%%%%%%%%%%%%%%%%%
\section{Conclusions and future work}
%%%%%%%%%%%%%%%%%%%%%%%%%%%%%%%%%%%%%%%%%%%%%%%%%%%%%%%%%%%%%%%%%%%%%%%%%%%%
\label{sec:Conclusion}
SOC is a key factor in soil health, and understanding its behavior and temporal and spatial distribution will be a key future research topic in remote science. Against this backdrop, we have developed an application (WALGREEN) that allows users from different backgrounds to load, manage, and visualize multispectral images and SOC values on a spatial and temporal basis. The tool also supports simple regression methods for SOC inference and temporal data-filling methods applied to the reflectance values from Sentinel-2 and Landsat.

WALGREEN helps democratize access to its data in real-time, regardless of the user's technical expertise. It lets users replicate experiments by saving results to a server. The user experience is designed to be intuitive, with a focus on providing SOC data and making it easier to handle. Users are abstracted from the technical integration processes underlying the front-end.

Future work to be done includes the integration of more advanced and complex regression methods, such as GPR~\cite{Rasmussen2006Gaussian} and data assimilation methods, including non-linear Kalman filtering~\cite{Sarkka2023}, as part of the ongoing development of this application.

%%%%%%%%%%%%%%%%%%%%%%%%%%%%%%%%%%%%%%%%%%%%%%%%%%%%%%%%%%%%%%%%%%%%%%%%%%%%
\section{Credit authorship contribution statement}
%%%%%%%%%%%%%%%%%%%%%%%%%%%%%%%%%%%%%%%%%%%%%%%%%%%%%%%%%%%%%%%%%%%%%%%%%%%%
\label{sec:authorship}
J. M. Aroca: conceptualization, data curation, software, visualization, writing – original draft; J. F. Díez: conceptualization, writing – original draft, project administration, resources; P. L. Carmona: conceptualization, writing – original draft, project administration, resources; V. Elvira: conceptualization, writing – original draft; G. Camps-Valls: conceptualization, writing – original draft; R. Pascual: writing – original draft; C. G. Osorio: software, visualization, writing – original draft.

%%%%%%%%%%%%%%%%%%%%%%%%%%%%%%%%%%%%%%%%%%%%%%%%%%%%%%%%%%%%%%%%%%%%%%%%%%%%
\section{Declaration of competing interest}
%%%%%%%%%%%%%%%%%%%%%%%%%%%%%%%%%%%%%%%%%%%%%%%%%%%%%%%%%%%%%%%%%%%%%%%%%%%%
\label{sec:sample:appendix}

The authors declare that they have no competing interests.

%%%%%%%%%%%%%%%%%%%%%%%%%%%%%%%%%%%%%%%%%%%%%%%%%%%%%%%%%%%%%%%%%%%%%%%%%%%%
\section*{Acknowledgments}
%%%%%%%%%%%%%%%%%%%%%%%%%%%%%%%%%%%%%%%%%%%%%%%%%%%%%%%%%%%%%%%%%%%%%%%%%%%%
\label{sec:Acknowledgments}
J. M. Aroca, J. F. Díez, P. L. Carmona, and C. G. Osorio acknowledge the financial support from
TED2021-131638B-I00, the ``Sentinel imagery to monitor
agricultural practices and their contribution to the initiative
`4 per 1000' increase in soil organic carbon''
(SEN4CFARMING) project of the Spanish Ministry of
Science and Innovation, and from the Recovery and Resilience
Facility of the European Commission.

\bibliographystyle{elsarticle-num}
\bibliography{bibliography/bibirefB}




\end{document}